\documentclass[ reprint, amsmath,amssymb, aps]{revtex4-2}

\usepackage{graphicx}
\usepackage{dcolumn}
\usepackage{bm}
\usepackage{amsmath}
\usepackage{amssymb}
\usepackage[usenames,dvipsnames]{xcolor} 
\usepackage{mathtools}
\usepackage[colorlinks=true,citecolor=blue,linkcolor=blue]{hyperref}
\usepackage[caption=false]{subfig}
\usepackage{bbm}
\usepackage{braket}
\usepackage{comment}
\usepackage{hyperref}
\usepackage{upgreek}
\usepackage{esint}
\usepackage{soul}
\usepackage[normalem]{ ulem } 
\usepackage{floatrow}

\begin{document}

\title{Anyon statistics through conductance measurements of time-domain interferometry}

\author{Noam Schiller$^1$}
\author{Yotam Shapira$^2$}
\author{Ady Stern$^1$}
\author{Yuval Oreg$^1$}
\affiliation{\small{$^1$Department of Condensed Matter Physics\\
$^2$Department of Physics of Complex Systems\\
Weizmann Institute of Science, Rehovot 7610001, Israel}}

\begin{abstract}
We propose a method to extract the mutual exchange statistics of the anyonic excitations of a general Abelian fractional quantum Hall state, by comparing the tunneling characteristics of a quantum point contact in two different experimental conditions. In the first, the tunneling current between two edges at different chemical potentials is measured. In the second, one of these edges is strongly diluted by an earlier point contact. We describe  the case of the dilute beam in terms of a time-domain interferometer between the  anyons flowing along the edge and quasiparticle-quasihole excitations created at the tunneling quantum point contact. In both cases, temperature is kept large, such that the measured current is given to linear response. Remarkably, our proposal does not require the measurement of current correlations, and allows us to carefully separate effects of the fractional charge and statistics from effects of intra- and inter-edge interactions.  
\end{abstract}
                              
\maketitle

\textit{Introduction.}---
It has been almost four decades since the initial proposal that the elementary quasiparticles of fractional quantum Hall (FQH) systems obey anyonic statistics \cite{arovas_fractional_1984}. Despite the apparent maturity of the field, the pursuit to definitively observe the physical quantities and quantum numbers characterizing anyons \cite{wen_quantum_2004,stern_anyons_2008} is constantly being reinvigorated \cite{laughlin_anomalous_1983,de-picciotto_direct_1997,saminadayar_observation_1997,kane_nonequilibrium_1994,kane_quantized_1997,griffiths_evolution_2000,kim_signatures_2005,stern_proposed_2006,bonderson_detecting_2006,vishveshwara_correlations_2010,campagnano_hanbury_2012,campagnano_chirality_2016,kim_measuring_2006,banerjee_observed_2017,banerjee_observation_2018,lee_negative_2019,Halperin_Jain_Book_2020}. In particular, early 2020 saw two major experimental steps forward: the observation of anyonic braiding in a Fabry-Perot interferometer \cite{nakamura_direct_2020}, and demonstration of a so-called ``anyon collider" \cite{bartolomei_fractional_2020,rosenow_current_2016} using cross-correlation measurements. 

\floatsetup[figure]{style=plain,subcapbesideposition=top}
\begin{figure}[h!]
    \centering
    \centering
    \sidesubfloat[]{
        \includegraphics[width=0.95 \columnwidth]
            {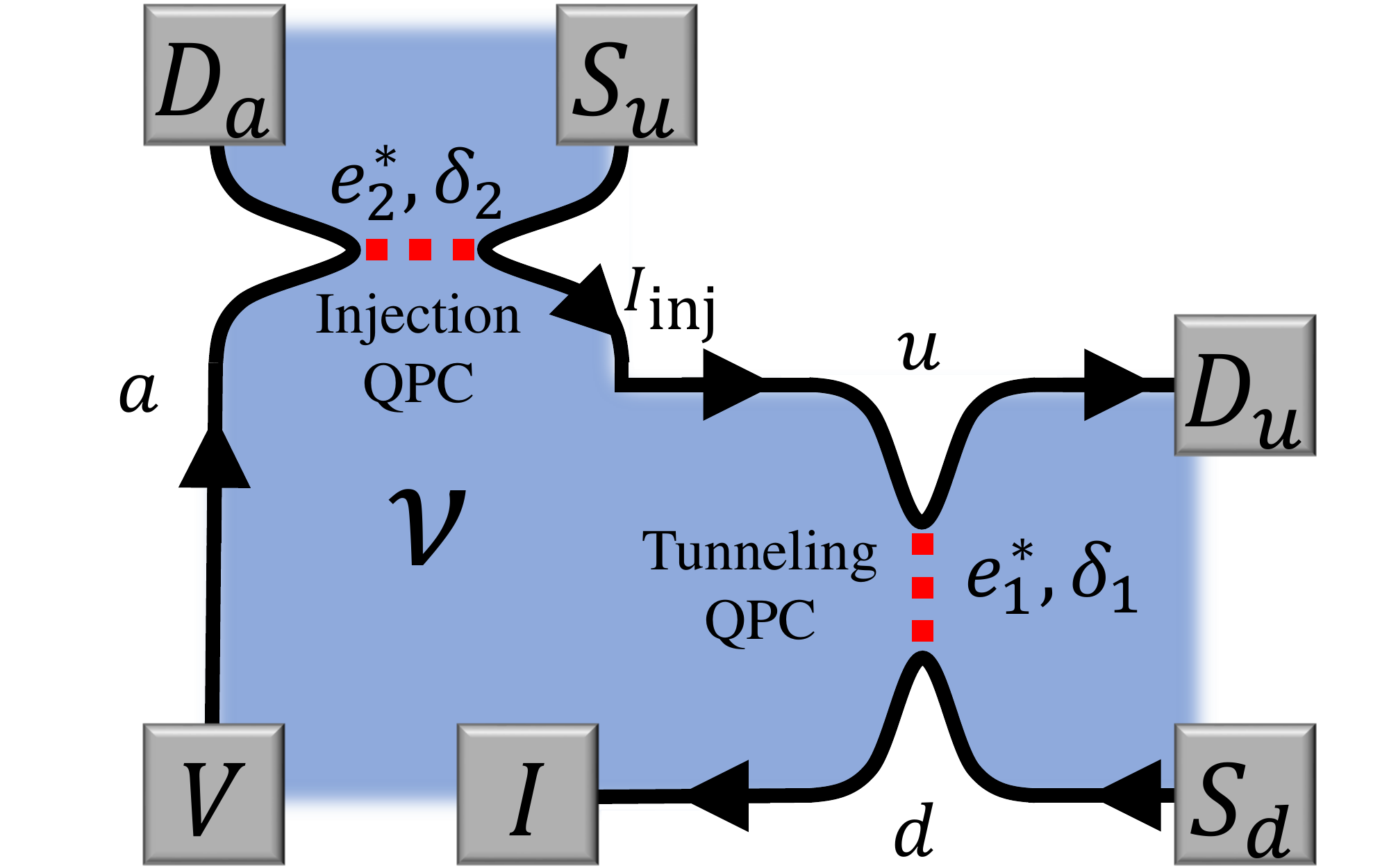}
        \label{subfig:Model}
    }
        \\[\baselineskip]
    \sidesubfloat[]{
        \includegraphics[width=\columnwidth]
            {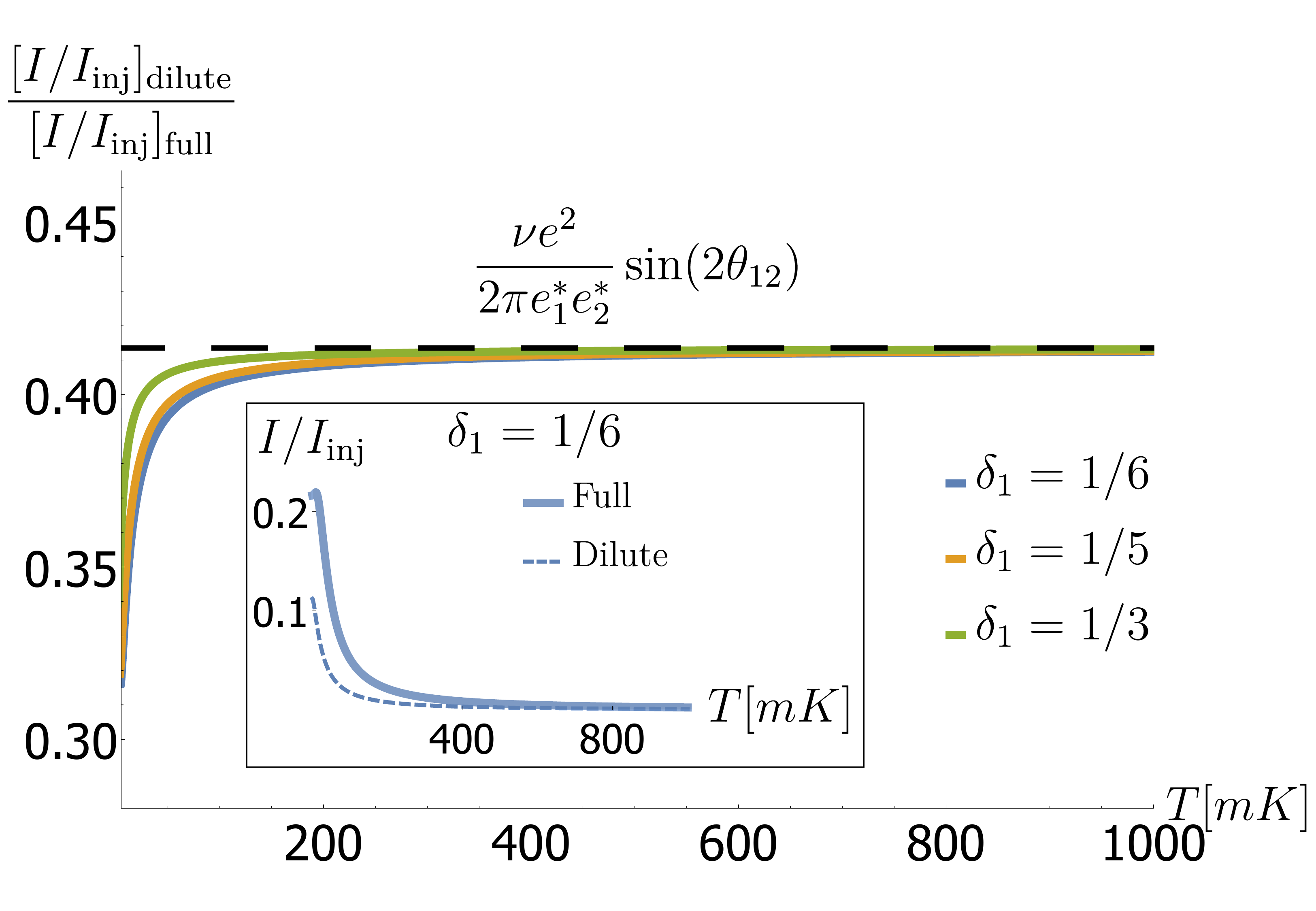}
        \label{subfig:Plot}
    }
    \caption{(a) Two counter-propagating edge modes ($u/d$) of a fractional quantum Hall droplet at filling factor $\nu$ are connected by a quantum point contact, through which quasiparticles of charge $e^*_1$ and scaling dimension $\delta_1$ can tunnel. Current is measured at the lower edge's drain, denoted by $I$. A current of $I_{\mathrm{inj}}$ is injected into the upper edge via a second, injection QPC, e.g. from a third auxiliary edge mode ($a$). The injection QPC is placed at a bias voltage of $V$, and allows tunneling of quasiparticles of charge $e^*_2$ and scaling dimension $\delta_2$. All other sources and drains are grounded. (b) The ratio between $I/I_{\mathrm{inj}}$ in the dilute case and $I/I_{\mathrm{inj}}$ in the full case, as a function of temperature, for $\nu=e^*_1/e=e^*_2/e=1/3$, and for different scaling dimensions $\delta_1$. For the dilute case, we $I_\mathrm{inj}= 10 \mathrm{pA}$, and assume $k_B T \ll e V$ for all relevant temperatures, such that the contribution from $G_\mathrm{direct}$ to Eq.~\eqref{eq:main_result} is negligible. In the full case, we use $V = 10 \upmu V$. Both cases use $\xi = 72\textrm{mK}, \tau_c = 10^{-13}\mathrm{s}$. When the dilute case satisfies $\hbar I_\mathrm{inj}/e \ll k_B T \ll eV \ll \hbar/\tau_c$, and the full case satisfies $\hbar I_\mathrm{inj}/e = \nu e V/2\pi \ll k_B T \ll \hbar/\tau_c$, the ratio approaches an asymptote that does not depend on scaling dimension, allowing extraction of the mutual statistics $\theta_{12}$. Inset: $I/I_{\mathrm{inj}}$ for the dilute and full cases as a function of temperature for $\delta_1 = 1/6$, the canonical value for a Laughlin $1/3$ state.}
    \label{fig:System}
\end{figure}

Here we show that anyonic statistics can be inferred directly from conductance measurements, without requiring current correlation measurements or explicitly building an interferometer. The configuration we propose to obtain this result consists of a quantum point contact (QPC) between two edges of a general Abelian FQH state which are driven out of equilibrium. The edges may be driven off-equilibrium by one of three methods: injecting a single quasiparticle into one of the edges; injecting a Poissonian, dilute beam of quasiparticles into one of the edges; and placing a finite bias voltage between the edges.

Our proposed setup, shown in Fig.~\hyperref[fig:System]{\ref{fig:System}(a)}, allows a smooth transition between the dilute Poissonian beam and a full beam at finite bias voltage. This is obtained by tuning a second, injection QPC from fully open (a differential conductance, $G_{\mathrm{inj}} \equiv d I_{\rm inj}/dV$, satisfying $G_\mathrm{inj}/\sigma_{xy} \rightarrow 0$) to fully closed ($G_{\mathrm{inj}}/\sigma_{xy} \rightarrow 1$). We henceforth refer to these as the dilute and full limits, respectively.

We propose sweeping $G_{\mathrm{inj}}$ through this range, and measuring the ratio $I/I_\text{inj}$, where $I$ is the measured current after the tunneling QPC, and $I_\text{inj}$ is the injected incident current, as defined in Fig.~\hyperref[fig:System]{\ref{fig:System}(a)}. Comparing the values at the dilute and full limits cancels out non-universal constants, yielding the relation, 
\begin{equation}
    \label{eq:main_result}
    \left[\frac{ I(T)}{I_{\rm inj}(T)} \right]_{\mathrm{dilute}}\!\!\!\!\!\!\!= 
    \frac{\nu e^2}{2 \pi e^*_1 e^*_2} \sin{2 \theta_{12}} \left[\frac{I(T)}{I_{\rm inj}(T)}\right]_{\mathrm{full}} \!\!\!\!\!+ \frac{G_\mathrm{direct}}{G_\mathrm{inj}}.
\end{equation}
Here, $e^*_1$ is the tunneling quasiparticle charge, $e^*_2$ the injected quasiparticle charge, $\delta_1$ is the tunneling quasiparticle scaling dimension, $\theta_{12}$ is the mutual statistics phase between the injected and tunneling quasiparticles, $T$ is temperature, and $G_\mathrm{direct}$ is a residual conductance corresponding to direct tunneling \cite{morel_fractionalization_2022,lee_non-abelian_2022,lee_partitioning_2022} through both QPCs. The full crossover between these two limits is shown schematically in Fig.~\hyperref[fig:System]{\ref{fig:System}(b)}.

\begin{figure}[t!]
    \centering
    \includegraphics[width=\columnwidth]{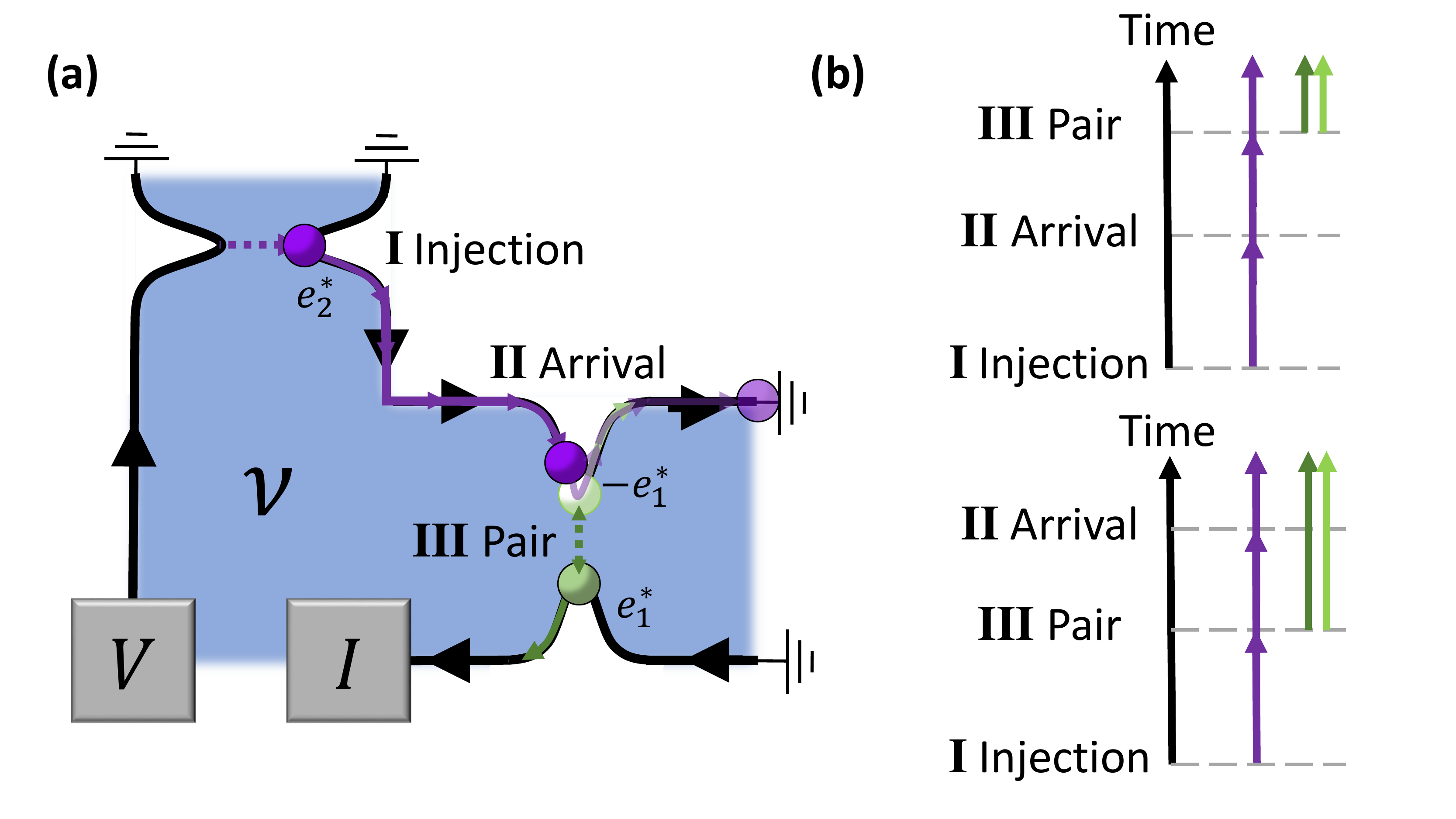}
    \caption{Time-domain interferometry. (a) I A quasiparticle is injected from the sourced, left edge, through the injection QPC, and into the upper edge. II The injected quasiparticle, by virtue of its chiral motion along the edge, arrives at the tunneling QPC. III A quasiparticle-quasihole pair is created at the tunneling QPC. (b) The two processes by which charge carriers may ultimately arrive at the drain. The injected quasiparticle arrives at the tunneling QPC either before (upper panel) or after (lower panel) the creation quasiparticle-quasihole pair. These two processes interfere, with a relative phase dictated by the mutual statistics phase, $e^{i 2 \theta_{12}}$.}
    \label{fig:interference}
\end{figure}

The mechanism leading to this result is a time-domain interferometer at the tunneling QPC which is created by the dilute incident beam. The interference is between two processes, in which a quasiparticle-quasihole excitation occurs at the tunneling QPC either before or after the arrival of an injected quasiparticle (see Fig.~\hyperref[fig:System]{\ref{fig:interference}}). A similar physical picture has been shown in Refs. \cite{han_topological_2016,lee_fractional_2020,lee_non-abelian_2022}. We further find that this interference is sensitive to the \textit{mutual statistics} phase between the injected and the tunneling quasiparticles, $\theta_{12}$. We emphasize that these quasiparticles are not necessarily of the same type, although they must be supported by the same FQH liquid.

Since our focus is the interference of two amplitudes which differ from one another by the orderings of events, the key point of our analysis is the identification of the phase differences between the two orderings. We find phase differences that are determined by the quasiparticle charge~$e^*$, which is a fraction of the electron charge for non-integer values of~$\nu$ \cite{laughlin_anomalous_1983,de-picciotto_direct_1997,saminadayar_observation_1997}; the scaling dimension~$\delta$, which defines the zero-temperature time correlations of the quasiparticle via the relation $\langle \psi^\dagger (\tau) \psi (0) \rangle \sim \tau ^{-2 \delta}$ \cite{yang_influence_2013,snizhko_scaling_2015,schiller_extracting_2022,ebisu_fluctuations_2022}; and the exchange statistics phase $\theta$, which for anyons take special values beyond the fermionic $\pi$ and the bosonic $2\pi$ \cite{arovas_fractional_1984,wen_quantum_2004,stern_anyons_2008}. 

We are interested here in isolating the effect of $\theta$ from the other two effects. In particular, we would like to separate it from the effect of $\delta$. For non-interacting edges, in which all the modes propagate in the same direction, $2 \pi \delta = \theta$; however, in general $\delta$ is affected by non-universal factors, such as intra-edge and inter-edge interactions, $1/f$ noise or neutral modes \cite{kane_transport_1992,kane_transmission_1992,rosenow_nonuniversal_2002,papa_interactions_2004,ferraro_relevance_2008,braggio_environmental_2012}. This in stark contrast to the charge, exchange statistics phase, or filling factor, which are universal.

We separate the effect of $\theta$ from that of $\delta$ by tuning the system to a regime where $\delta$ only affects observables through a non-universal prefactor, which then cancels out in the ratio of currents given in Eq.~\eqref{eq:main_result}. We arrive at this regime by employing a careful ordering of the various energy scales in the system, such that $\hbar I_{\mathrm{inj}}/e \ll k_B T$ throughout the entire crossover of $G_{\mathrm{inj}}$. This ensures that the current $I$ is given to linear response in $I_{\mathrm{inj}}$. We present an analytic expression generalizing Eq.~\eqref{eq:main_result} outside of this regime in App~\ref{app:Full_derivation}, Eq.~\eqref{eq:general_results}.

While in the full limit the edge that enters the tunneling QPC is in equilibrium at chemical potential $V$, at the dilute limit we need the injection QPC to reflect only a small fraction of the impinging electrons, such that the resulting injection current is Poissonian and rare. Said differently,  the injected current in this limit must satisfy $I_{\mathrm{inj}} \ll \sigma_{xy} V$. Furthermore, the beam must still be dilute when arriving at the tunneling QPC. As such, the distance between the two QPCs must be sufficiently small that no equilibration or dephasing occurs along the way. Finally, we assume that tuning the injection QPC does not affect the transparency of the tunneling QPC, to ensure that all non-universal constants are cancelled when examining the ratio of the two limits. \footnote{In practice this may be difficult to implement experimentally, as the tuning of both QPCs is most easily done through gating. This obstacle may be overcome by performing the dilute beam measurements as shown in  Fig.~\hyperref[fig:System]{\ref{fig:System}(a)}, while performing the full beam measurements by biasing the lower edge through source $S_d$, and measuring the tunneling current to the upper edge at drain $D_u$.}

Easy extraction of $\theta_{12}$ requires $G_\mathrm{direct}$ to be sub-dominant (see Eq.~\eqref{eq:main_result}). Quantitatively, this is the case if both $k_B T \ll e V$ and $4 \delta_1 < 2$ are satisfied. These constraints result from the direct tunneling process being dominated by short time scales. Naive theories describing quasiparticles may satisfy this condition even if the aforementioned non-universal effects change the scaling dimension quite significantly. For example, theory gives $\delta = 1/2m$ for Laughlin quasiparticles.

\textit{Edge theory.}---
We now define the system's Hamiltonian and derive the current. As shown by Wen, the edge theory of a general Abelian FQH state can be described by $n$-boson fields, $\boldsymbol{\phi} (x,t) \equiv \left( \phi_1, \phi_2, \cdots \phi_n \right)^{T}$ \cite{wen_quantum_2004}. These define the theory in conjunction with a charge vector, $\boldsymbol{q}$, which determines the electric charge carried by each boson field, and the so called $K$-matrix, which determines the commutation relations between the boson fields,
\begin{equation}
    \left[\phi_i(x),\partial_{x^\prime}\phi_j(x^\prime)\right] = i 2\pi (K ^{-1} )_{ij} \delta (x-x^\prime).
    \label{commrelation}
\end{equation} 
The filling factor is then given by $\nu= \boldsymbol{q}^T K^{-1} \boldsymbol{q}$, and the charge density is given by $\rho = -\frac{1}{2\pi}\boldsymbol{q}\cdot \partial_x \boldsymbol{\phi}$. In terms of these fields, the Hamiltonian of a single FQH edge mode is given by
\begin{equation}
    \label{eq:single_edge}
    \mathcal{H_{\rm edge}}= \frac{1}{4\pi} \sum_{i,j=1}^n \int dx \partial_x \phi_i V_{ij} \partial_x \phi_j,
\end{equation}
where $\hat{V}$ is a positive definite matrix describing the velocities of the modes and intra-edge interactions. These edges support quasiparticles of the form $\psi_{\boldsymbol{l}}\sim e^{i \boldsymbol{l}\cdot \boldsymbol{\phi}}$, where $\boldsymbol{l}$ is a vector of integers. The charge of these quasiparticles is then given by $e^*_{\boldsymbol{l}}= \boldsymbol{q}^T K^{-1} \boldsymbol{l}$.

The configuration of Fig.~\hyperref[fig:System]{\ref{fig:System}(a)} involves two edges, $u$ and $d$, tunnel-coupled by a QPC. This is described by two copies of the Hamiltonian $\mathcal{H_{\rm edge}}$, time reversed with regard to one another, as well as a tunneling term, $\mathcal{H}_T$, which we treat as a perturbation. Assuming only one type of quasiparticle, denoted by the vector $\boldsymbol{l}_1$ and carrying charge $e^*_1$, tunnels between the edges, this is given by 
\begin{equation}
    \label{eq:tunneling}
    \mathcal{H}_T = \xi \left[\hat{A}+\hat{A}^\dagger \right];
    \;
    \hat{A}(t) \equiv e^{i \left(\boldsymbol{l}_1\cdot \boldsymbol{\phi}^{(u)}(0,t) - \boldsymbol{l}_1\cdot \boldsymbol{\phi}^{(d)}(0,t)\right)}.
\end{equation}
Here, $\xi$ is a small tunneling amplitude, which we assume to be real, and $\boldsymbol{\phi}^{(u)}$ ($\boldsymbol{\phi}^{(d)}$) are the bosonic field operators on the upper (lower) edge. We project the auxiliary edge $a$ out of the Hamiltonian, as it is only used to ``initialize" the state of the edge $u$. 

The current that tunnels from the upper edge to the lower edge is then given by the operator, $\hat{I}_T(t) = i \xi e^*_{1} \left[\hat{A}^\dagger(t)-\hat{A}(t)\right]$. Since the lower edge is grounded, we henceforth identify $I = \langle \hat{I}_T \rangle$. Expanding to leading order in $\xi$, the current is given by
\begin{equation}
    \label{eq:Observables}
        I(t) = e^*_1 \xi^2  \int_{-\infty}^{t}  dt^\prime 
        \left<  \left[ \hat{A}^\dagger(t),\hat{A}(t^\prime) \right] +  \left[ \hat{A}^\dagger(t^\prime),\hat{A}(t) \right] \right>.
\end{equation}
Here, $[\cdot,\cdot]$ denotes commutation, and expectation values are calculated with respect to the Hamiltonian in the absence of tunneling. 

\textit{Deviation from Equilibrium.}---
It is clear from Eq.~\eqref{eq:Observables} that one needs to derive correlation functions such as $\langle \hat{A}^\dagger(t)\hat{A}(t^\prime) \rangle$. In equilibrium, at temperature $T$, the system is particle-hole symmetric, and the correlation functions are given by \cite{wen_quantum_2004,giamarchi_quantum_2003}
\begin{align}
    \label{eq:correlation_equilibrium}
    \langle \hat{A}^\dagger(t)\hat{A}(t^\prime) \rangle_0 &= 
    \langle \hat{A}(t)\hat{A}^\dagger(t^\prime) \rangle_0  \\ &=
    \left[ \frac{\pi T \tau_c}{\sinh \left( \pi T \left|t-t^\prime \right| \right)}\right]^{4 \delta_{1}} e^{-i 2\pi \delta_1 \textrm{sgn}\left(t-t^\prime\right)},\nonumber
\end{align}
where $\delta_{1}$ is the scaling dimension of the quasiparticle $\boldsymbol{l}_1$, and $\tau_c >0$ is a short time cutoff. 

Two main features are carried over from Eq.~\eqref{eq:correlation_equilibrium} to the correlation functions out of equilibrium - the exponential decay at time difference larger than $\hbar / T$, and the phase $e^{2 \pi i \delta_1}$ associated with an interchange of the time arguments. 

We now consider two non-equibrium cases. In the first we introduce a constant bias voltage $V\equiv V_u -V_d$ between the edges. In the setup of Fig.~\hyperref[fig:System]{\ref{fig:System}(a)}, this corresponds to a fully closed injection QPC, i.e. $I_\mathrm{inj}=\sigma_{xy}V$. The introduction of the voltages can be formally absorbed into the boson fields by use of a simple gauge transformation, which maps $\boldsymbol{\phi}^{(u/d)}(x,t) \mapsto \boldsymbol{\phi}^{(u/d)}(x,t) + K^{-1} \boldsymbol{q} V_{u/d} \left( t\mp x/v \right)/\hbar$. This accordingly modifies the correlation functions by a phase factor
\begin{equation}
    \begin{aligned}
        \label{eq:correlation_voltage}
        \langle \hat{A}^\dagger(t)\hat{A}(t^\prime) \rangle_\mathrm{full} & = 
        \langle \hat{A}^\dagger(t)\hat{A}(t^\prime) \rangle_0 e^{i \frac{e^*_{1} V}{\hbar}(t-t^\prime)}, \\
        \langle \hat{A}(t)\hat{A}^\dagger(t^\prime) \rangle_\mathrm{full} & = 
        \langle \hat{A}(t)\hat{A}^\dagger(t^\prime) \rangle_0 e^{-i \frac{e^*_{1} V}{\hbar}(t-t^\prime)}. 
    \end{aligned}
\end{equation}

In the second non-equilibrium driving, we consider injecting a single quasiparticle, denoted by the vector $\boldsymbol{l}_2$, into the upper edge at the location $x_\mathrm{inj}<0$ and at time $t_\mathrm{inj}$. This is shown schematically in Fig.~\hyperref[fig:interference]{\ref{fig:interference}(a)}.  In view of the commutation relations \eqref{commrelation}, the application of the quasiparticle creation operator $e^{-i\boldsymbol{l}_2\cdot{\boldsymbol \phi^{(u)}(x_\mathrm{inj},t_\mathrm{inj})}}$ on  the edge creates a soliton in each of the boson fields, 
\begin{equation}
     \label{solitonshift}
    \boldsymbol{\phi}^{(u)}(x,t_\mathrm{inj}) \mapsto \boldsymbol{\phi}^{(u)} (x,t_\mathrm{inj})
    - 2 \pi K^{-1} \boldsymbol{l}_{2} \Theta \left(x - x_\mathrm{inj} \right).
\end{equation}
We assume here the injection happens instantaneously. This assumption will be relaxed to find the subleading term of Eq.~\eqref{eq:main_result}.

The fields at general times can then be obtained using the equations of motion dictated by the Hamiltonian in Eq.~\eqref{eq:single_edge}. If all modes are chiral with the same velocity $v$, this amounts to replacing $x-x_\mathrm{inj} \rightarrow x-x_\mathrm{inj} -v \left(t-t_\mathrm{inj} \right)$. The soliton thus arrives at the QPC, $x=0$, at time $t_0 \equiv t_\mathrm{inj}-x_\mathrm{inj}/v$.

The $c$-number shift in the bosonic field of Eq.~\eqref{solitonshift} leads to a phase shift in the correlator Eq.~\eqref{eq:correlation_equilibrium}. We see directly from the definition of the operator $\hat{A}$ in Eq.~\eqref{eq:tunneling} that
\begin{equation}
    \begin{aligned}
        \label{eq:Braiding_phase}
    \langle \hat{A}^\dagger(t)\hat{A}(t^\prime) \rangle_\mathrm{qp} & =
    \langle \hat{A}^\dagger(t)\hat{A}(t^\prime) \rangle_0  e^{ 2 \pi i \boldsymbol{l}_1K^{-1} \boldsymbol{l}_2 \left[ \Theta \left( t - t_0 \right) -\Theta \left( t^\prime - t_0 \right) \right] }, \\
    \langle \hat{A}(t)\hat{A}^\dagger(t^\prime) \rangle_\mathrm{qp} & =
    \langle \hat{A}(t)\hat{A}^\dagger(t^\prime) \rangle_0  e^{ -2 \pi i \boldsymbol{l}_1K^{-1} \boldsymbol{l}_2 \left[ \Theta \left( t - t_0 \right) -\Theta \left( t^\prime - t_0 \right) \right] }.
    \end{aligned}
\end{equation}

The phase we obtain is the standard definition of mutual braiding statistics between two quasiparticles, $\theta_{12} \equiv \pi \boldsymbol{l}_1K^{-1} \boldsymbol{l}_2$ \cite{wen_quantum_2004}. The expression in Eq. \eqref{eq:Braiding_phase} shows that the product gains a phase of $e^{2i\theta_{12} \mathrm{sgn}\left(t-t^\prime\right)}$ if the arrival time $t_0$ is between the times $t^\prime$ and $t$, and a trivial phase of 1 otherwise. We emphasize how naturally this result came from the underlying theory: the only assumptions necessary to obtain this are the commutation relations, \eqref{commrelation}, and the existence of quasiparticles in the edge's excitation spectrum.

This result holds for different boson modes with different velocities if all solitons arrive at the tunneling QPC more or less concurrently, avoiding dephasing. This is the case if $|x_\mathrm{inj}|/\Delta v \ll \hbar/ T$, where $\Delta v$ is the velocity difference between the fastest and the slowest modes.

\textit{Time-domain interferometry.}---
The appearance of the phase, $\theta_{12}$, can be understood as time-domain interferometry of the two distinct $\pm e^*_1$ quasiparticle-quasihole excitations, before and after the injected $e^*_2$ quasiparticle arrives at the QPC. A similar physical picture has been shown in Ref. \cite{han_topological_2016,lee_fractional_2020,lee_non-abelian_2022}.

To show this we consider the configuration of a single injected particle, as described in Fig.~\hyperref[fig:interference]{\ref{fig:interference}(a)}. In this case the non-equilibrium correlation function takes the form, 
\begin{equation}
    \langle \hat{A}^\dagger(t)\hat{A}(t^\prime) \rangle_\mathrm{qp}=\langle \psi_{\boldsymbol{l}_2}(t_0) \hat{A}^\dagger(t)\hat{A}(t^\prime) \psi^\dagger_{\boldsymbol{l}_2}(t_0) \rangle_0,\label{eq:noneq_c}
\end{equation}
i.e., the expectation value is calculated with respect to the state resulting from exciting the ground state $\ket{0}$ with a single quasiparticle. Here we omit the position variable from the quasiparticle injection operator $\psi^\dagger_{\boldsymbol{l}_2}(t_0)$, and assume it arrives at the tunneling QPC $x=0$ at time $t_0$.

The current in Eq.~\eqref{eq:Observables} is then given by integration over multiple terms of the form in Eq.~\eqref{eq:noneq_c}. We define $\Ket{t,t_0}_- \equiv \hat{A}(t) \psi^\dagger_{\boldsymbol{l}_2}(t_0) \ket{0}$ and $\Ket{t,t_0}_+ \equiv \hat{A}^\dagger(t) \psi^\dagger_{\boldsymbol{l}_2}(t_0) \ket{0}$. Eq.~\eqref{eq:Observables} can now be re-written as
\begin{equation}
    \label{eq:interference}
         I  \propto -\int_{-\infty}^t dt'\sum_{b=\pm} 
        b \big| \ket{t,t_0}_b+\ket{t',t_0}_b \big|^2.
\end{equation}

The expression above involves two interference terms. The term with $b=-$ is an interference between creation of $-e^*_1$ quasiholes on the upper edge at the QPC at times $t$ and $t^\prime$. The two interfering processes are shown schematically in Fig.~\hyperref[fig:interference]{\ref{fig:interference}(b)}. As shown in the first row of Eq.~\eqref{eq:Braiding_phase}, these two processes are distinguished by a non-trivial phase of $e^{i2\theta_{12}}$ if the arrival time $t_0$ is in between the quasiholes' creation times, $t'<t_0<t$. Combined with the equilibrium correlation function Eq.~\eqref{eq:correlation_equilibrium}, one finds that this interference gives a term proportional to $\cos \left( 2\theta_{12} - 2 \pi \delta \right)$.
Using similar arguments, the term with $b=+$ in Eq.~\eqref{eq:interference}, gives an interference term proportional to $\cos \left( 2\theta_{12} + 2 \pi \delta \right)$. The total contribution from the two terms in Eq.~\eqref{eq:interference} is thus proportional to $\sin \left( 2\theta_{12}\right) \sin \left( 2\pi \delta \right)$ \footnote{This would appear to give zero for $\delta = 1/2$; however, this cancels out with a divergence resulting from integration over the coordinate $t^\prime$.}.

This interference happens entirely in the time domain, and along only one edge. It is however crucial that this edge be part of a two-dimensional bulk. This is important both because the second edge is required to absorb the leftover quasiparticle or quasihole resulting from the pair creation at the QPC, and because the injected quasiparticle must be created within a bulk FQH droplet. Furthermore, the bulk is intimately related to the edge through bulk-edge correspondence. This dictates that the statistical phase contributing to time-domain interference along a single edge, which our setup measures, is the same as the phase obtained from spatial exchange.

It is easy to generalize this to injection of multiple quasiparticles: as long as all injected quasiparticles are mutually independent, each injected quasiparticle contributes a phase of $e^{2i\theta_{12}}$ if and only if the arrival time at the point contact was between $t^\prime$ and $t$. If we assume this is a Poissonian process, with a quasiparticle injection rate of $I_{\mathrm{inj}}/e^*_2$, we obtain for $t>0$
\begin{equation}
    \begin{aligned}
    \label{eq:Poissonian}
    \frac{\langle \hat{A}^\dagger(t)\hat{A}(0) \rangle_\mathrm{dilute}}{\langle \hat{A}^\dagger(t)\hat{A}(0) \rangle_0} 
    & = \sum_{n=0}^{\infty}  
    \frac{(t I_\mathrm{inj}/e_2^*) ^n e^{-t I_\mathrm{inj}/e_2^*}}{n!} e^{2in\theta_{12}} \\
    & = e^{-t I_\mathrm{inj}/e_2^* \left( 1- e^{2 i \theta_{12}}\right)}.
    \end{aligned}
\end{equation}
This is precisely the result given in Refs.~\cite{rosenow_current_2016,lee_non-abelian_2022} for injection along a single edge. Adding injected quasiparticles to the lower edge and generalizing for $t<0$ are straightforward using the same arguments.

\textit{Currents.}---
The effect of driving the system out of equilibrium is completely encapsulated in the correlation functions obtained above. These can then be used to derive any observable of interest, such as charge or heat currents in any of the system's drains, or their respective auto- and cross-correlations. 

For concreteness, we present the explicit results of such a calculation for the charge current at the lower drain, denoted as $I$ in Fig.~\ref{fig:System}. We show that a simple cohort of current measurements is sufficient to obtain the mutual statistics $\theta_{12}$, without requiring correlation measurements.

We focus on the regime where the temperature is large compared to the injected current $\hbar I_{\mathrm{inj}}/e k_B T$. For the full limit, this assumption guarantees linear response in the voltage and in the injected current, which in this limit is $I_\mathrm{inj} = \sigma_{xy} V$. For the dilute limit, the exponential suppression of the equilibrium correlation function at times larger than $\hbar/T$, guarantees that the exponent in Eq. (\ref{eq:Poissonian}) may be expanded to first order in $I_{\mathrm{inj}}$. Consequently, 
\begin{equation}
    \label{eq:linear_response}
     \frac{\langle \hat{A}^\dagger(t)\hat{A}(t^\prime) \rangle_\mathrm{full/dilute}}{\langle \hat{A}^\dagger(t)\hat{A}(t^\prime) \rangle_0} 
     \approx 1 + i \omega_\mathrm{f/d} \left(t - t^\prime \right),
\end{equation}
where the frequencies $\omega_\mathrm{f/d}$ are given by
\begin{equation}
    \omega_\mathrm{f} = \frac{e^*_1 V}{\hbar} = \frac{e^*_1}{\hbar} \frac{I_\mathrm{inj}}{\sigma_{xy}}; \quad
    \omega_\mathrm{d} = i\frac{I_\mathrm{inj}}{e^*_2} \left( 1- e^{2i\theta_{12}} \right).
\end{equation}

The zeroth order term corresponds to the equilibrium state and does not contribute to the current. The ratio of the two first order contributions is Eq. (\ref{eq:main_result}). 

Explicit calculation of the resulting current in Eq.~\eqref{eq:Observables}, given in App.~\ref{app:Full_derivation}, finds that
\begin{equation}
    \label{eq:linear_response_current}
    I_\mathrm{full/dilute} = 2 \pi e^*_1 (\xi \tau_c)^2 (2 \pi T \tau_c)^{4 \delta_1 -2} \mathcal{B} \left( 2 \delta_1 ,2 \delta_1 \right) \mathrm{Re} \left[ \omega_\mathrm{f/d} \right],
\end{equation}
where $\mathcal{B}(x,y)$ is the Euler Beta function. It is thus immediately apparent that by focusing on the ratio between the full and dilute beams, all dependence on $\delta_1, T$ and $\xi$ drops out. Examining the ratio $I/I_\mathrm{inj}$, and noting that $\sigma_{xy} \hbar / e^*_1 e^*_2 =  \nu e^2/ 2\pi e^*_1 e^*_2 $ we thus obtain Eq.~\eqref{eq:main_result}.

For general temperatures, the current can no longer be treated as a linear response to the drive of the full or dilute beams. We hence obtain the typical power laws characterizing tunneling in Luttinger liquids \cite{wen_edge_1991,kane_transmission_1992,chamon_tunneling_1995,wen_quantum_2004}. Comparing measurements of the full and dilute limits at the low temperature limit $T \ll e^* V, I_{\mathrm{inj}}$ can still give a quantity related to the mutual statistics $\theta_{12}$, but will explicitly depend on the value of $\delta_1$. We present general expressions for the current in this case in App.~\ref{app:Full_derivation}.

For a fermionic $\theta_{12} = \pi$, Eq.~\eqref{eq:linear_response_current} gives no current at all for a dilute electron beam. However,  Landauer-Buttiker-Imry scattering theory \cite{blanter_shot_2000} tells us the current is given by the product of the transparencies of the two QPCs along the electron's path, regardless of whether they are close to full transmission or full reflection. This requires accounting for the direct tunneling term in Eq.~\eqref{eq:main_result}, which now becomes the leading contribution.

We do this by accounting for the finite width of the soliton. This leads to the required, Landauer-Buttiker-Imry consistent result of $I_{\mathrm{dilute}} = 4 \pi ^2 \tau_c^2 \xi^2 I_{\mathrm{inj}}$. The physical intuition behind the requirement of a finite soliton width is that tunneling without time-domain interferometry, dubbed the direct tunneling process in \cite{lee_non-abelian_2022,morel_fractionalization_2022}, is dominated by short times. Performing these calculations explicitly in App.~\ref{app:Solitons}, we show that the ratio between the first term in Eq.~\eqref{eq:main_result} and $G_\mathrm{direct}$ is $ \propto \left( T \tau_s\right)^{4\delta_1 -2}$, where $\tau_s$ is the soliton width. It has been shown \cite{lee_non-abelian_2022,morel_fractionalization_2022} that $\tau_s^{-1} \propto \max \{eV,k_B T\}$; as such, to ensure $G_\mathrm{direct}$ is sub-dominant, the dilute limit must be measured when $k_B T \ll eV$ and $4 \delta_1 < 2$.

Several contemporary experimental setups use the equivalent of non-interacting fermionic formulae to reasonable success \cite{feldman_why_2017}, corresponding to the limiting value of $2 \delta_1 =1$. In this case, the second term of Eq.~\eqref{eq:main_result} is a numerical coefficient of order one, which may depend solely on $e^*, \delta_1$ and $\theta_{12}$. For non-interacting fermions, this coefficient is easily found by comparing to known Landauer-Buttiker-Imry scattering theory \cite{blanter_shot_2000}, but it is straightforward to generalize. We discuss this coefficient further in App.~\ref{app:Solitons}.

\textit{Discussion.}---
We propose a simple method to extract anyonic exchange statistics. Our system consists only of a single quantum Hall droplet with two QPCs, which effectively create a time-domain interferometer, as can be identified from current measurements. We thus avoid both current correlation (or noise) measurements, and the need for a real space interferometer, making the identification of the exchange statistics much more accessible than existing experiments. All time-domain interferometry is between pairs of an injected quasiparticle and a tunneling quasiparticle, and occurs at the same edge, as previously proposed in Ref.~\cite{lee_non-abelian_2022}. 

Both the exchange statistics $\theta_{11}$ of the tunneling quasiparticle, and $\theta_{22}$ of the injection quasiparticle, do not appear in our derivation. Rather, it is the two particles' \textit{mutual statistics}, $\theta_{12}$ that affect the modified correlation functions, and hence, the physical observables. Likewise, the scaling dimension and electric charge which directly effect observables are only those of the tunneling quasiparticle, $\delta_{1}$ and $e^*_1$ (properties of the injected quasiparticles may implicitly enter through the injection rate).

Only in the case where the injected and tunneling quasiparticles are identical, $\boldsymbol{l}_1=\boldsymbol{l}_2$, do we obtain exchange statistics for a single quasiparticle type. We remark that this is indeed the case in the experiment of Ref.~\cite{bartolomei_fractional_2020}, where all quasiparticles are Laughlin $e^*=e/3$ anyons, and subsequent recreations for the $\nu=1/3$ and $\nu=2/5$ cases \cite{lee_partitioning_2022,ruelle_comparing_2022,glidic_cross-correlation_2022}. Interestingly, a recent experiment employing a similar setup, where the injected quasiparticle was a $e/3$ anyon and the tunneling quasiparticle was an electron, observed Andreev-like reflection \cite{glidic_quasiparticle_2022}. This is consistent with a mutual statistics phase of $\theta_{12}=\pi$, for which Eq.~\eqref{eq:main_result} gives no time-domain interferometry signal.

\textit{Acknowledgements.}--- We thank Tomer Alkalay, Moty Heiblum, Changki Hong, June-Young Lee and H.-S. Sim for insightful discussions and comments on the manuscript. This work was partially supported by grants from the ERC
under the European Union’s Horizon 2020 research and innovation programme
(grant agreements LEGOTOP No. 788715 and HQMAT No. 817799), the DFG
(CRC/Transregio 183, EI 519/7-1), the BSF and NSF (2018643), the ISF
Quantum Science and Technology (2074/19). N.S. was supported by the Clore Scholars Programme.

\bibliography{main}

\newpage

\appendix
\begin{widetext}
\setcounter{page}{1}
\setcounter{table}{0}
\renewcommand\thetable{\Alph{section}.\arabic{table}}

\section{Finite temperature current from time-domain interferometry}
\label{app:Full_derivation}

Here derive explicit expressions for the tunneling current $I$ at finite temperature $T$. This section neglects the contribution $G_\mathrm{direct}$ (see Eq.~\eqref{eq:main_result}, which is discussed in App.~\ref{app:Solitons}. We begin with the expression for the current in Eq.~\eqref{eq:Observables}. Writing this explicitly,
\begin{equation}
    \label{eq:Explicit_current}
    I = e^*_1 \xi^2  \int_{-\infty}^{t}  dt^\prime \bigg \{
        \left<  \hat{A}^\dagger(t)\hat{A}(t^\prime) \right> - 
        \left<  \hat{A}(t^\prime)\hat{A}^\dagger(t) \right> +
        \left<  \hat{A}^\dagger(t^\prime)\hat{A}(t) \right> - 
        \left<  \hat{A}(t)\hat{A}^\dagger(t^\prime) \right>
        \bigg \}.
\end{equation}
In the case where the edges are not driven out of equilibrium, we plug the equilibrium correlation functions Eq.~\eqref{eq:correlation_equilibrium}, and obtain $I=0$, as expected. A similar expression can be written for the symmetrized current fluctuations, 
\begin{equation}
    \label{eq:fluctuations}
      \left<  \Big\{ \delta\hat{I}_T(t),  \delta\hat{I}_T(t^\prime) \Big\} \right>  = (e^*_1)^2 \xi^2  
        \bigg \{
        \left<  \hat{A}^\dagger(t)\hat{A}(t^\prime) \right> + 
        \left<  \hat{A}(t^\prime)\hat{A}^\dagger(t) \right> +
        \left<  \hat{A}^\dagger(t^\prime)\hat{A}(t) \right> + 
        \left<  \hat{A}(t)\hat{A}^\dagger(t^\prime) \right>
        \bigg \},
\end{equation}
where we define $\delta\hat{I}_T \equiv \delta\hat{I}_T - \langle \delta\hat{I}_T \rangle$. We do not focus on current fluctuations in this work, but note that our methods reproduce the known results of Refs.~\cite{rosenow_current_2016,lee_non-abelian_2022}.

We now want to obtain the current for each of the three methods of driving the two edges out of equilibrium. Each of these leads to a corresponding multiplicative factor to the correlation functions. A finite bias voltage $V$, used for the ``full" beam, gives the correlation functions of Eq.~\eqref{eq:correlation_voltage}; injection of a single quasiparticle gives the correlation functions of Eq.~\eqref{eq:Braiding_phase}; and a dilute, Poissonian beam of quasiparticles gives the correlation functions of Eq.~\eqref{eq:Poissonian}. Plugging in these appropriate correlation functions gives after minor algebra and changes of variables
\begin{subequations}
    \begin{align}
        \label{eq:Full_Derivation_All}
        I_\mathrm{full} & = 2 i e^*_1 \xi^2  \int_{0}^{\infty}  d \tilde{t} 
        \sin \left(\frac{e^*_1 V}{\hbar} \tilde{t}  \right)
        \biggr \{ \left[ \frac{\pi T \tau_c}{i \sinh \left( \pi T \left(\tilde{t}-i\tau_c \right) \right)}\right]^{4 \delta_{1}} - \left[ \frac{\pi T \tau_c}{i \sinh \left( \pi T \left(-\tilde{t}-i\tau_c \right) \right)}\right]^{4 \delta_{1}} \biggr \}, \\
        I_\mathrm{dilute} & = 2i e^*_1 \xi^2  
        \int_{0}^{\infty}  d \tilde{t}
        \frac{\sin \left( \frac{I_\mathrm{inj}}{e^*_2} \tilde{t} \sin{2 \theta_{12}}\right)}
        {\exp \left( \frac{I_\mathrm{inj}}{e^*_2} \tilde{t} \left(1-\cos{2 \theta_{12}}\right)\right)}
        \biggr \{ \left[ \frac{\pi T \tau_c}{i \sinh \left( \pi T \left(\tilde{t}-i\tau_c \right) \right)}\right]^{4 \delta_{1}} - \left[ \frac{\pi T \tau_c}{i \sinh \left( \pi T \left(-\tilde{t}-i\tau_c \right) \right)}\right]^{4 \delta_{1}} \biggr \},
        \\
        I_\mathrm{qp} & = 2i e^*_1 \xi^2  \int_{-\infty}^{t}  dt^\prime 
        \sin \left( 2 \theta_{12} \left[ \Theta \left( t - t_0 \right) -\Theta \left( t^\prime - t_0 \right) \right] \right)
        \biggr \{ \left[ \frac{\pi T \tau_c}{i \sinh \left( \pi T \left(t-t^\prime-i\tau_c \right) \right)}\right]^{4 \delta_{1}} - \left[ \frac{\pi T \tau_c}{i \sinh \left( \pi T \left(t^\prime-t-i\tau_c \right) \right)}\right]^{4 \delta_{1}} \biggr \}.
    \end{align}
\end{subequations}

We proceed using the identity, correct at the limit $\tau_c \rightarrow 0$,
\begin{equation}
    \label{eq:Dirac_Identity}
    i \biggr \{ \left[ \frac{\pi T \tau_c}{i \sinh \left( \pi T \left(\tilde{t}-i\tau_c \right) \right)}\right]^{4 \delta_{1}} - \left[ \frac{\pi T \tau_c}{i \sinh \left( \pi T \left(-\tilde{t}-i\tau_c \right) \right)}\right]^{4 \delta_{1}} \biggr \} =
    \begin{cases}
    - 2 \pi \tau_c^2 \partial_{\tilde{t}}\delta(\tilde{t}) & 2 \delta_1 = 1 \\
     \left[ \frac{\pi T \tau_c}{\sinh \left( \pi T |\tilde{t}|\right)}\right]^{4 \delta_{1}} 2 \sin{(2 \pi \delta_1)} \mathrm{sgn}(\tilde{t}) & 2 \delta_1 \neq 1
    \end{cases},
\end{equation}
where $\delta(t)$ is the Dirac delta function. This identity is necessary to treat the case of $\delta_1 = 1$, which otherwise may lead to divergent integrals. 

Standard manipulations of these integrals then give results in terms of the Euler Beta function and the incomplete Beta function, $\mathcal{B}\left( x; a,b\right) \equiv \int_0^x t^{a-1}(1-t)^{b-1}dt$, $\mathcal{B}\left(a,b\right)\equiv \mathcal{B}\left( 1; a,b\right)$. We thus obtain the general results
\begin{subequations}
    \label{eq:general_results}
    \begin{align}
        I_\mathrm{full} & = 2 e^*_1 \xi^2 (2 \pi T)^{4 \delta_1 -1} \tau_c^{4\delta_1} \sinh \left(\frac{e^*_1 V}{2 T}\right) \mathcal{B} \left( 2 \delta_1 +i \frac{e^*_1 V}{2 \pi  T},2 \delta_1 -i \frac{e^*_1 V}{2 \pi T} \right) \\
        I_\mathrm{dilute} & = -\frac{2 \pi}{\cos \left(2 \pi \delta_1 \right) \Gamma \left(4 \delta_1 \right)} e^*_1 \xi^2 (2 \pi T)^{4 \delta_1 -1} \tau_c^{4 \delta_1} \mathrm{Im} \left[ 
        \frac{ \Gamma \left( \frac{I_\mathrm{inj}}{e^*_2} \frac{1-\cos \left(2 \theta_{12} \right) + i \sin \left(2 \theta_{12} \right)}{2 \pi T} + 2 \delta_1\right) }{\Gamma \left( \frac{I_\mathrm{inj}}{e^*_2} \frac{1-\cos \left(2 \theta_{12} \right) + i \sin \left(2 \theta_{12} \right)}{2 \pi T} + 1-2 \delta_1\right)}
        \right]
        \\
        I_\mathrm{qp} & =  4 e^*_1 \xi^2 (2 \pi T)^{4 \delta_1 -1} \tau_c^{4\delta_1} \sin \left( 2\theta_{12}\right) \sin \left( 2 \pi \delta \right) \mathcal{B}\left( e^{-2 \pi T \left(t-t_0 \right)} ; 1+2 \delta_1, 1 - 4 \delta_1 
    \right).
    \end{align}
\end{subequations}
Here, $\Gamma(a)$ is the Euler Gamma function, satisfying $\mathcal{B}(a,b)=\frac{\Gamma (a) \Gamma (b)}{\Gamma (a+b)}$, and $\mathrm{Im}[\dots]$ denotes the imaginary part.

The high temperature and zero temperature limits of the full beam and dilute beam are then immediately reproducible. For $e^* V, \hbar I_\mathrm{inj}/e^*_2 \ll k_BT$, one expands to leading order in $e^*V/T$ and $I_\mathrm{inj}/e^*_2 T$, one obtains
\begin{subequations}
    \label{eq:high_temp}
    \begin{align}
         I_\mathrm{full} & \approx 2 \pi e^*_1 (\xi \tau_c)^2 (2 \pi T \tau_c)^{4 \delta_1 -2}  \mathcal{B} \left( 2 \delta_1 ,2 \delta_1 \right) \frac{e^*_1 V}{\hbar} , \\
         I_\mathrm{dilute} & \approx 2 \pi  e^*_1 (\xi \tau_c)^2 (2 \pi T \tau_c)^{4 \delta_1 -2}   \mathcal{B} \left( 2 \delta_1 ,2 \delta_1 \right) \frac{I_\mathrm{inj}}{e^*_2} \sin \left(2 \theta_{12}\right).
    \end{align}
\end{subequations}
We thus see that the mutual statistics are immediately extractable from the dilute case. While this expression does depend on the non-universal $\xi$ and $\delta$, as well as the temperature, these are all encoded in a prefactor which appears in the full case as well. We can hence lose this unwanted prefactor by examining the ratio between the two cases.

For $T \ll e^* V, I_\mathrm{inj}/e^*_2$, we use the identities
\begin{align*}
    \lim_{x\rightarrow\infty} \Gamma \left(x + a \right) & = \Gamma \left(x \right) x^a, \\
    \lim_{x\rightarrow\infty} \sinh (\pi x) \mathcal{B} \left( a+ix,a-ix \right) & = \frac{\pi}{\Gamma (2a)} x^{2a-1},
\end{align*}
to obtain 
\begin{subequations}
    \begin{align}
    \label{eq:zero_temp}
         I_\mathrm{full} & \approx \frac{2 \pi e^*_1 \xi^2 \tau_c^{4\delta_1}}{\Gamma \left(4 \delta_1\right)} \left( \frac{e^*_{1} V}{\hbar} \right)^{4 \delta_1 -1}, \\
         I_\mathrm{dilute} & \approx 
    - \frac{2 \pi e^*_1 \xi^2\tau_c^{4 \delta_1}}
    {\cos{\left(2 \pi \delta_1 \right)} \Gamma \left(4 \delta_1\right)}
     \left( \frac{I_{\mathrm{inj}}}{e^*_2}\right)^{4\delta_1-1} \mathrm{Im} 
    \bigg[ \bigg( 1-\cos{\left( 2\theta_{12} \right)} + i \sin{\left( 2 \theta_{12} \right)}\bigg)^{4\delta_1 -1} \bigg].
    \end{align}
\end{subequations}
By tuning $2 \delta_1 \rightarrow 1$, we again obtain an expression from which the mutual statistics are easily extractable, with an identical non-universal prefactor appearing in both the full and dilute cases. However, once the scaling dimension is tuned to this critical value, the contribution from time-domain interferometry no longer dominates the direct tunneling process, as can be seen from the calculation of $G_\mathrm{direct}$ in App.~\ref{app:Solitons}.

We note that for temperatures larger than the source voltage, one has to account for injection of both quasiparticles and quasiholes through the injection QPC. This can be done by modifying the Poissonian correlation function in Eq.~\eqref{eq:Poissonian} according to
\begin{equation}
    \begin{aligned}
    \label{eq:Poissonian modified}
    \frac{\langle \hat{A}^\dagger(t)\hat{A}(0) \rangle_\mathrm{dilute}}{\langle \hat{A}^\dagger(t)\hat{A}(0) \rangle_0} 
    & = e^{-t I_\mathrm{inj}/e_2^* \left( 1- e^{2 i \theta_{12}}\right)} \\
    & \rightarrow e^{-t I_\mathrm{inj}^{qp}/e_2^* \left( 1- e^{2 i \theta_{12}}\right)} e^{-t I_\mathrm{inj}^{qh}/e_2^* \left( 1- e^{-2 i \theta_{12}}\right)},
    \end{aligned}
\end{equation}
where $I_\mathrm{inj}^{qp}$ is the injection rate of quasiparticles, and $I_\mathrm{inj}^{qh}$ is the injection rate of quasiholes. This is a similar expression to the three QPC setup considered in \cite{rosenow_current_2016} and \cite{lee_non-abelian_2022}. Performing the same algebra as in this section, and identifying $I_\mathrm{inj} \equiv I_\mathrm{inj}^{qp} - I_\mathrm{inj}^{qh}$, one then reproduces Eq.~\eqref{eq:high_temp} for the high temperature limit.

Finally, it is instructive to consider the current due to the injection of a single quasiparticle at time $t_0$, which was obtained in Eq.~\hyperref[eq:general_results]{(\ref{eq:general_results}c)}. In this case we must examine the explicit temperature dependence, as tunneling of a single quasiparticle may be relevant, and we lack any other energy scale to serve as a cutoff for the RG flow of the process. This current exhibits a power-law decay for $t-t_0 \ll 1/\pi T$, consistent with the orthogonality catastrophe that characterizes injection into Luttinger liquid edges. For $2 \delta_1=1$, this results in $\langle \hat{I}_T \rangle_{\mathrm{qp}} \propto \delta \left(t-t_0\right)$. This gives some intuition as to what makes the $2 \delta_1 = 1$ case so unique - the QPC just scatters the incident particle with some probability, without inducing any long-time correlations, resulting in the direct tunneling process.

\section{Finite soliton width: restoring Landauer-Buttiker-Imry for electrons and subleading corrections}
\label{app:Solitons}

The results of App.~\ref{app:Full_derivation} are seemingly inconsistent with the known non-interacting electron limits. Indeed, inserting $e^*_1=e^*_2=e$, $2 \delta_1=2 \delta_2=1$ and $\theta_{12} = \pi$ into these results would indicate that the dilute electron beam gives no current at all. This is in direct contrast with the intuition of Landauer-Buttiker-Imry scattering theory, which would indicate that the current should be given by the product of the transparencies of the two QPCs along the electron's path, regardless of whether they are close to full transmission or full reflection.

The culprit of this result is a peculiarity of soliton physics. The boson field $\phi$ is compact under $\phi \mapsto \phi + 2\pi$. As such, a soliton of height $2 \pi K^{-1} \boldsymbol{l}_2$ would appear to leave the boson field completely unperturbed if $K^{-1} \boldsymbol{l}_2$ is an integer. This corresponds precisely to electron injection operators \cite{wen_quantum_2004}. As such, our soliton description is ill-equipped to treat electrons without modifications.

We solve this issue by introducing a finite width to the soliton, $\tau_s$. To fully recreate the known non-interacting result, it is crucial to maintain an order of limits such that the soliton width is larger than the short-time cutoff, $\tau_c$. We note that we still take care to ensure that $\tau_s < 1/T, (I_\mathrm{inj}/e^*_2)^{-1}$, i.e. the solitons are still narrow compared to the larger time scales in the problem. Previous works \cite{lee_non-abelian_2022,morel_fractionalization_2022}, performing a full Keldysh calculation, have shown the soliton width (refered to in the cited papers as the temporal width) is given by the voltage, $h/e^*V$, if $eV > k_B T$, and by the inverse temperature $\hbar/k_B T$ if $eV \lesssim k_B T$; as such, the dilute limit must be measured in the regime $ I_\mathrm{inj}/e^*_2 \ll k_B T \ll eV$.

Formally, this means that injecting a quasiparticle into the upper edge at the location $x_0$ and time $t_0$ transforms the boson field according to
\begin{equation*}
    \boldsymbol{\phi}^{(u)}(x,t_0) \mapsto \boldsymbol{\phi}^{(u)} (x,t_0)
    - 2 \pi K^{-1} \boldsymbol{l}_{2} \left( \frac{1}{\pi} \tan^{-1}\left( \frac{x-x_0}{\tau_s}\right) - \frac{1}{2}\right).
\end{equation*}
Accordingly, the correlation functions of Eq.~\eqref{eq:Braiding_phase} are now replaced with
\begin{equation}
    \begin{aligned}
        \label{eq:Braiding_phase_soliton}
    \langle \hat{A}^\dagger(t)\hat{A}(t^\prime) \rangle_\mathrm{qp} & =
    \langle \hat{A}^\dagger(t)\hat{A}(t^\prime) \rangle_0  
    \exp{ \left(2 i \frac{\theta_{12}}{\pi} \left[ \tan^{-1} \left( \frac{t - t_0}{\tau_s} \right) -\tan^{-1} \left( \frac{t^\prime - t_0}{\tau_s} \right) \right] \right)}, \\
    \langle \hat{A}(t)\hat{A}^\dagger(t^\prime) \rangle_\mathrm{qp} & =
    \langle \hat{A}(t)\hat{A}^\dagger(t^\prime) \rangle_0  
    \exp{ \left(-2 i \frac{\theta_{12}}{\pi} \left[ \tan^{-1} \left( \frac{t - t_0}{\tau_s} \right) -\tan^{-1} \left( \frac{t^\prime - t_0}{\tau_s} \right) \right] \right)}.
    \end{aligned}
\end{equation}
One indeed sees that at the limit $\tau_c \rightarrow 0$, one reproduces the immediate soliton results from the main text.

To find the correlation function in the presence of a dilute, Poissonian beam of injected quasiparticles, we now must sum over the number of injected quasiparticles, in a manner similar to Eq.~\eqref{eq:Poissonian}. However, this is now trickier, for two reasons. First, the accumulated phase explicitly depends on the time of the injected quasiparticle. Second, injected quasiparticles outside of the window $[0,t]$ can still affect the correlation function, due to the long tails of the finite-width solitons.  

So generalizing the methods that lead to Eq.~\eqref{eq:Poissonian}, the correlation function now changes to define
\begin{equation}
    \label{eq:Full_Poissonian_Correlation}
    \begin{aligned}
    \frac{\langle \hat{A}^\dagger(t)\hat{A}(0) \rangle_\mathrm{fw}}{\langle \hat{A}^\dagger(t)\hat{A}(0) \rangle_0} 
    & = \sum_n  
    \frac{\left((t+2c \tau_c )\frac{I_\mathrm{inj}}{e_2^*} \right) ^n e^{-(t+2 c \tau_c) \frac{I_\mathrm{inj}}{e_2^*}}}{n!} 
    \left[ \int_{-c\tau_c}^{t+c\tau_c} d t_0 P \left( \textrm{Particle injected at } t_0\right) 
    e^{ 2 i \frac{\theta_{12}}{\pi} \left[ \tan^{-1} \left( \frac{t - t_0}{\tau_s} \right) -\tan^{-1} \left( \frac{0 - t_0}{\tau_s} \right) \right] }\right]^n.
    \end{aligned}
\end{equation}
Here $c$ is some unitless cutoff, chosen such that injected quasiparticles affect the correlation function only if they are injected in the window $[-c\tau_c,t+c \tau_c]$, which we will eventually take to be infinite. The probability of injection at a particular time $t_0$ is given by
\begin{equation}
    \label{eq:Poissonian_distribution_t0}
    P \left( \textrm{Particle injected at }t_0\right) = \frac{I_\mathrm{inj}/e^*_2 e^{-I_\mathrm{inj}  t_0/e^*_2}}{\int_{-c\tau_c}^{t+c\tau_c} dt_0I_\mathrm{inj}/e^*_2 e^{-I_\mathrm{inj}  t_0/e^*_2}}. 
\end{equation}
Performing this sum, and re-defining this integration with unitless variables, we find that the new correlation function is given in integral form by
\begin{equation}
    \label{eq:Poissonian_finite_width}
    \begin{aligned}
    \frac{\langle \hat{A}^\dagger(t)\hat{A}(0) \rangle_\mathrm{fw}}{\langle \hat{A}^\dagger(t)\hat{A}(0) \rangle_0} 
    & = \exp{ \left( -\left(t+ 2 c \tau_c\right) 
    \frac{I_\mathrm{inj}}{e^*_2}
    \left[ 1 - \mathcal{I}_{\theta_{12}} \left( \frac{I_\mathrm{inj}}{e^*_2} \tau_s,\frac{t}{2 \tau_s}\right)\right]\right)}, \\
    \mathcal{I}_{\theta_{12}} \left( a,b\right) & \equiv
    \frac{a}{2 \sinh \left(a(b+c)\right)} \int_{-b-c}^{b+c}dx e^{-a x} e^{2 i \frac{\theta_{12}}{\pi}\left[\tan^{-1}\left(x+b\right)-\tan^{-1}\left(x-b\right)\right]}.
    \end{aligned}
\end{equation}
By plugging this new correlation function into the expression for the current in Eq.~\eqref{eq:Observables}, one now finds
\begin{multline}
    \label{eq:Dilute_with_Itheta}
     I_\mathrm{dilute} = 2i e^*_1 \xi^2  
        \int_{0}^{\infty}  d \tilde{t}
        \frac{\sin \left( (\tilde{t}+2c\tau_c) \frac{I_\mathrm{inj}}{e^*_2} \mathrm{Im} \left[\mathcal{I}_{\theta_{12}} \left( \frac{I_\mathrm{inj}}{e^*_2}\tau_s,\frac{t}{2 \tau_s}\right) \right]\right)}
        {\exp \left( (\tilde{t}+2c\tau_c) \frac{I_\mathrm{inj}}{e^*_2} \mathrm{Re} \left[1-\mathcal{I}_{\theta_{12}} \left( \frac{I_\mathrm{inj}}{e^*_2}\tau_s,\frac{t}{2 \tau_s}\right) \right]\right)} \\
        \times \biggr \{ \left[ \frac{\pi T \tau_c}{i \sinh \left( \pi T \left(\tilde{t}-i\tau_c \right) \right)}\right]^{4 \delta_{1}} - \left[ \frac{\pi T \tau_c}{i \sinh \left( \pi T \left(-\tilde{t}-i\tau_c \right) \right)}\right]^{4 \delta_{1}} \biggr \}.
\end{multline}
Careful re-application of the limit $\tau_c \rightarrow 0$ indeed replicates our previous result in Eq.~\eqref{eq:Poissonian}.

For general $\theta_{12}$, the integral $ \mathcal{I}_{\theta_{12}} \left( a,b\right)$ is difficult to solve analytically. In the main text, this is circumvented by taking the limit $\tau_c \rightarrow 0$, allowing use of Eq.~\eqref{eq:Dirac_Identity}, in conjunction with replacing $(\tilde{t}+2c\tau_c) \frac{I_\mathrm{inj}}{e^*_2} \mathcal{I}_{\theta_{12}} \left( \frac{I_\mathrm{inj}}{e^*_2}\tau_s,\frac{t}{2 \tau_s}\right) \rightarrow - i \tilde{t} \omega_d$. However, as noted previously, fermionic exchange statistics corresponding to values of $\theta_{12}$ that are integer multiples of $\pi$ lead to $\omega_d =0$, and hence give a vanishing current. As such, Eq.~\eqref{eq:Dilute_with_Itheta} must be calculated in full while retaining a finite $\tau_c$. 

To simplify these expressions,  we assume that $\left(\frac{I_\mathrm{inj}}{e^*_2} \right)$ is significantly larger than any other time scale in the system. This makes sense from a physical perspective as well, as it corresponds to the assumption that injection is sufficiently rare such that solitons do not overlap. In this case, one can assume the probability of injection which appears in Eqs.~\eqref{eq:Full_Poissonian_Correlation},\eqref{eq:Poissonian_distribution_t0} is approximately uniform, i.e. $P \left( \textrm{Particle injected at }t_0\right) \approx 1/(t+2c\tau_c)$. One can now safely take the limit $c \rightarrow \infty$ without artificial divergences, giving the simpler result,
\begin{equation}
    \label{eq:finite_soliton_integral}
     \frac{\langle \hat{A}^\dagger(t)\hat{A}(0) \rangle_\mathrm{fw}}{\langle \hat{A}^\dagger(t)\hat{A}(0) \rangle_0} 
    = \exp{ \left( 
    \frac{I_\mathrm{inj}}{e^*_2}
    \int_{-\infty}^{\infty} dt_0 
    \left[     e^{ 2 i \frac{\theta_{12}}{\pi} \left[ \tan^{-1} \left( \frac{t - t_0}{\tau_s} \right) -\tan^{-1} \left( \frac{0 - t_0}{\tau_s} \right) \right] } -1 \right]\right)}.
\end{equation}

Since we undertook this endeavor with the explicit goal of finding the correct result for non-interacting electrons, we wish to find this integral for $2 \delta_1 = 1$, $\theta_{12} = \pi$, and $e^*_1=e^*_2=e$. This value of $\theta_{12}$ allows one to significantly simplify Eq.~\eqref{eq:finite_soliton_integral} using trignometric identities; plugging the resulting correlation function in Eq.~\eqref{eq:Explicit_current}, we obtain
\begin{equation}
    \label{eq:Finite_Soliton_Current_theta_pi}
    I_\mathrm{dilute}^{\theta_{12}=\pi} = 2i e^*_1 \xi^2  
        \int_{0}^{\infty}  d \tilde{t}
        \frac{\sin \left( \frac{I_\mathrm{inj}}{e^*_2}  
        \frac{2 \pi \tilde{t} (2 \tau_s)^2}{\tilde{t}^2 + (2 \tau_s)^2} 
        \right)}
        {\exp \left(\frac{I_\mathrm{inj}}{e^*_2}  
        \frac{2 \pi \tilde{t}^2 (2 \tau_s)}{\tilde{t}^2 + (2 \tau_s)^2} \right)}
        \biggr \{ \left[ \frac{\pi T \tau_c}{i \sinh \left( \pi T \left(\tilde{t}-i\tau_c \right) \right)}\right]^{4 \delta_{1}} - \left[ \frac{\pi T \tau_c}{i \sinh \left( \pi T \left(-\tilde{t}-i\tau_c \right) \right)}\right]^{4 \delta_{1}} \biggr \}.
\end{equation}

As can be seen in Eq.~\eqref{eq:Dirac_Identity}, the expression in the curled brackets is approximately zero for $\tilde{t}>\tau_c$. We can thus approximate the total integral as the contribution from short times, $\tilde{t} \leq \tau_c \ll 1/\pi T$. To leading order, this will be given by
\begin{equation}
    \label{eq:restore_LB}
    \begin{aligned}
    I_\mathrm{dilute}^{\theta_{12}=\pi,2 \delta_1 = 1} & \approx 2i e^*_1 \xi^2 \tau_c ^2  
        \int_{0}^{\infty}  d \tilde{t}
        \frac{I_\mathrm{inj}}{e^*_2}  
        \frac{2 \pi \tilde{t} (2 \tau_s)^2}{\tilde{t}^2 + (2 \tau_s)^2} 
        \biggr \{ \left( \frac{1}{i\tilde{t}+\tau_c }\right)^2 - \left( \frac{1}{-i\tilde{t}+\tau_c }\right)^2 \biggr \} \\
        & = \frac{(2\tau_s)^2}{(2\tau_s+\tau_c)^2}4\pi^2 \xi^2 \tau_c ^2 I_\mathrm{inj}.
    \end{aligned}
\end{equation}

Now taking the limit $\tau_c \ll \tau_s$, we compare to the electron case in, say, Eq.~\eqref{eq:linear_response_current} or Eq.~\eqref{eq:general_results}. We find that the result we expect for non-interacting electrons is indeed $4 \pi^2 \xi^2 \tau_c ^2 I_\mathrm{inj}$. This is consistent with  - the current is linear in the injected current, and in the transparency of the tunneling QPC (which is given by $\xi ^2 \tau_c ^2$). 

For general values of $\theta_{12}$ and $\delta_1$ this integral is more difficult to solve analytically. However, it is possible to re-write Eq.~\eqref{eq:finite_soliton_integral} as

\begin{align}
    \frac{\langle \hat{A}^\dagger(t)\hat{A}(0) \rangle_\mathrm{fw}}{\langle \hat{A}^\dagger(t)\hat{A}(0) \rangle_0} &= \exp{
    \frac{I_\mathrm{inj}}{e^*_2}
    \bigg( \sin \left( 2 \theta_{12} \right)t + f_{\theta_{12}}\left(t,\tau_c)\right) \bigg)}, \\
    f_{\theta_{12}}\left(t,\tau_c)\right) &\propto
    \begin{cases}
      t  & t \lesssim \tau_s \\
      \tau_s & t \gg \tau_s, \theta_{12} \neq \pi  \\
      (\tau_s)^2/t & t \gg \tau_s, \theta_{12} = \pi.
    \end{cases}
\end{align}

Plugging this into the general expression for the current, and expanding to linear response in $\frac{I_\mathrm{inj}}{e^*_2}$ we find 

\begin{equation}
    \label{eq:Finite_Soliton_general}
    I_\mathrm{dilute} = 2i e^*_1 \xi^2 \frac{I_\mathrm{inj}}{e^*_2}  
        \int_{0}^{\infty}  d \tilde{t}
        \bigg( \sin \left( 2 \theta_{12} \right)t + f_{\theta_{12}}\left(t,\tau_c)\right) \bigg)
        \biggr \{ \left[ \frac{\pi Thw \tau_c}{i \sinh \left( \pi T \left(\tilde{t}-i\tau_c \right) \right)}\right]^{4 \delta_{1}} - \left[ \frac{\pi T \tau_c}{i \sinh \left( \pi T \left(-\tilde{t}-i\tau_c \right) \right)}\right]^{4 \delta_{1}} \biggr \}.
\end{equation}
The term proportional to $\sin \left( 2 \theta_{12} \right)$, as discussed at length above, is the main interest of this paper. This is calculated in Eq.~\eqref{eq:high_temp}. We see there that the time scales in the system contribute a leading term of the form $\propto (\xi \tau_c)^2 (T \tau_c)^{4\delta_1 -2} I_\mathrm{inj} $. 

The term proportional to $f_{\theta_{12}}\left(t,\tau_c)\right)$ contains several contributions: at short times ($\tilde{t}\sim \tau_c$), we obtain a contribution of order $(\xi \tau_c) ^2 $; at long times ($\tilde{t}\sim 1/\pi T$) we obtain a contribution of order $(\xi \tau_c) ^2 (\tau_s/\tau_c)\left( T \tau_c \right)^{4\delta_1-1} $ for $\theta_{12}\neq \pi$ and $(\xi \tau_c) ^2 (\tau_s/\tau_c)\left( T \tau_c \right)^{4\delta_1} $ for $\theta_{12} = \pi$; and at intermediate times ($\tilde{t}\sim \tau_s$) we obtain contributions of order $(\xi \tau_c) ^2 (\tau_s/\tau_c)^{1-4\delta_1} $ and $(\xi \tau_c) ^2 (\tau_s/\tau_c)^{2-4\delta_1} $. 

We compare these contributions to the coefficients of Eq.~\eqref{eq:linear_response_current} or Eq.~\eqref{eq:high_temp}, which give the time-domain interferometry process, which is of order $(\xi \tau_c) ^2 \left( T \tau_c \right)^{4\delta_1-2} $. Utilizing $\tau_c \ll \tau_s \ll 1/\pi T$, we see that the long time contribution is always subdominant, but the short time dominates for $2 \delta_1 \geq 1$ - consistent with both Eq.~\eqref{eq:restore_LB} and the known electron result. This is consistent with our physical intuition: direct tunneling dominates short times, which give the main contribution for $2\delta_1 \geq 1$, whereas time-domain interferometry dominates long times, which give the main contribution for $2\delta_1 < 1$. 

Finally, if we indeed assume $2\delta_1<1$, the intermediate time contribution dominates the entire direct process. In this case, the ratio between the time-domain interferometry process and the direct process is given by $\propto(T \tau_s)^{4\delta_1 - 2}$. This again confirms that we must have a soliton width smaller than the inverse temperature to ensure time-domain interferometry

This method is also what we use to calculate the current for an almost full beam, i.e. $\sigma_{xy}-G_\mathrm{inj} \ll 1$. Since in this case, the beam can be treated as a conjoined full beam of fractional quasiparticles with a dilute beam of $e^* = e$ holes, we have $2 \theta_{12} = 2 \pi n$ regardless of the tunneling quasiparticles. Defining the injection rate of holes as $I_\mathrm{inj}^{\mathrm{holes}} \equiv \sigma_{xy} V - I_\mathrm{inj}$, we combine the full beam correlation function of Eq.~\eqref{eq:correlation_voltage} and the regularized Poissonian hole injection to obtain described in this section
\begin{multline}
    I_\mathrm{|G_\mathrm{inj}-\sigma_{xy}|\ll1} = 2i e^*_1 \xi^2  
        \int_{0}^{\infty}  d \tilde{t}
        \frac{\sin \left( \frac{e^*_1 V}{\hbar }\tilde{t} -\frac{I_\mathrm{inj}^{\mathrm{holes}}}{e} \frac{2 \pi \tilde{t} (2 \tau_s)^2}{\tilde{t}^2 + (2 \tau_s)^2}\right)}
        {\exp \left(\frac{I_\mathrm{inj}^{\mathrm{holes}}}{e} \frac{2 \pi \tilde{t}^2 (2 \tau_s)}{\tilde{t}^2 + (2 \tau_s)^2}\right)} \\
        \times\biggr \{ \left[ \frac{\pi T \tau_c}{i \sinh \left( \pi T \left(\tilde{t}-i\tau_c \right) \right)}\right]^{4 \delta_{1}} - \left[ \frac{\pi T \tau_c}{i \sinh \left( \pi T \left(-\tilde{t}-i\tau_c \right) \right)}\right]^{4 \delta_{1}} \biggr \}.
\end{multline}

In the relevant limits, the same methods as previously mention allow us to approximate the exponent in the denominator as $1$, and to expand the sine in the numerator. We thus have the sum of two linear responses, one in in $\frac{e^*_1 V}{\hbar}$ and one in $- \frac{I_\mathrm{inj}^{\mathrm{holes}}}{e}$. Taking, as in the Landauer-Buttiker-Imry case, the limit $\tau_s \gg \tau_c$, i.e. a soliton width that is larger than the short time cutoff, this can be re-written as

\begin{equation}
    I_\mathrm{|G_\mathrm{inj}-\sigma_{xy}|\ll1} \approx  2i e^*_1 \xi^2  
        \int_{0}^{\infty}  d \tilde{t}
        \left(\frac{e^*_1 V}{\hbar } -2\pi \frac{I_\mathrm{inj}^{\mathrm{holes}}}{e} \right)\tilde{t} \biggr \{ \left[ \frac{\pi T \tau_c}{i \sinh \left( \pi T \left(\tilde{t}-i\tau_c \right) \right)}\right]^{4 \delta_{1}} - \left[ \frac{\pi T \tau_c}{i \sinh \left( \pi T \left(-\tilde{t}-i\tau_c \right) \right)}\right]^{4 \delta_{1}} \biggr \}.
\end{equation}
Identifying $\left(\frac{e^*_1 V}{\hbar } -2\pi \frac{I_\mathrm{inj}^{\mathrm{holes}}}{e} \right) = \frac{2 \pi}{e}\left(\sigma_{xy}V-I_\mathrm{inj}^{\mathrm{holes}}\right) \equiv \frac{2 \pi}{e} I_\mathrm{inj}$, we see that this is precisely the same integral that we had in Eq.~\eqref{eq:Full_Derivation_All} for the full beam case, with the replacement $\sigma_{xy} V \rightarrow  I_\mathrm{inj}$. We note that we used here $\sigma_{xy} = e e^*/h$, which is correct only for Laughlin edge states, $\nu = 1/m$; this is valid as Laughlin edges are the outer level of heirarchal FQH fluids, and thus are the states of interest for nearly full closed QPCs.

\end{widetext}
\end{document}